%% file: CMSinvH.tex
\documentclass[12pt]{article}
\usepackage{graphicx}
\usepackage{slashed}
\usepackage{amsmath}
\usepackage{color}
\usepackage{colortbl}
\usepackage{xspace}
\definecolor{darkgreen}{cmyk}{1,0,1,0.2}
\def\METL{\slashed{E}_\mathrm{T}}
\def\MET{\ETslash}
\def\RMET{\text{red-}\ETslash}

\def\ZZexpl{\text{ZZ $\rightarrow \ell^+ \ell^- \cPgn \cPagn$}}
\def\ZHexpl{\text{ZH $\rightarrow \ell^+ \ell^- + \text{H}($inv$)$}}

\newcommand\pubnumber{DPF2013-307} 
\newcommand\pubdate{\today}

\def\neu{Department of Physics, Northeastern University, Boston, MA, 02115, USA}

\def\Title#1{\begin{center} {\Large #1 } \end{center}}
\def\Author#1{\begin{center}{ \sc #1} \end{center}}
\def\Address#1{\begin{center}{ \it #1} \end{center}}

\newcommand\pubblock{\rightline{\begin{tabular}{l} \pubnumber\\
         \pubdate  \end{tabular}}}
\newenvironment{Abstract}{\begin{quotation}  }{\end{quotation}}
\newenvironment{Presented}{\begin{quotation} \begin{center} 
             PRESENTED AT\end{center}\bigskip 
      \begin{center}\begin{large}}{\end{large}\end{center} \end{quotation}}

\input ptdr-definitions.sty

\input econfmacros.tex

\begin{document}
\begin{titlepage}
\pubblock

\vfill
\Title{Search for invisible Higgs boson production with the CMS detector at the LHC}
\vfill
\Author{Matthew E. Chasco for the CMS Collaboration}
\Address{\neu}
\vfill
\begin{Abstract}
Results are presented for the search for invisible Higgs boson production using the full LHC dataset corresponding to integrated luminosity of 5.1 $\fbinv$ and 19.6 $\fbinv$ of proton-proton collision data at $\sqrt{s}=$ 7 TeV and 8 TeV (respectively) collected by the CMS detector. 
The invisible Higgs is searched for in final states of missing transverse energy, with two leptons from a recoiling Z boson. 
No significant excess is found beyond standard model predictions, and limits are obtained on the branching fraction of the Higgs boson to invisible particles.
\end{Abstract}
\vfill
\begin{Presented}
DPF 2013\\
The Meeting of the American Physical Society\\
Division of Particles and Fields\\
Santa Cruz, California, August 13--17, 2013\\
\end{Presented}
\vfill
\end{titlepage}
\def\thefootnote{\fnsymbol{footnote}}
\setcounter{footnote}{0}

\section{Introduction}
The observation of a Higgs boson at the LHC has prompted studies to search for any connections this boson may have to new physics.
If a significant invisible branching fraction of the Higgs boson is observed, it would strongly suggest physics beyond the standard model.
Both CMS and ATLAS have set indirect constraints on invisible Higgs boson decays using measured rates of visible decay modes.
The current limits on the invisible branching fraction are 64\% for CMS indirect search~\cite{CMSind}, 60\% for ATLAS indirect search~\cite{ATLASind}, and 65\% for ALTAS direct search~\cite{ATLASdir}.
This analysis covers a direct search, done at CMS, for invisible Higgs decays, with full 7 and 8 TeV datasets.
We concentrate on the Higgs mass of 125 GeV, but also extend the search for other possible Higgs bosons in the mass range of 105-145 GeV.
\section{Experimental Apparatus}

The Compact Muon Solenoid (CMS)~\cite{CMS} comprises several subdetector systems, each constructed to detect a particular set of particle types and related quantities.
The primary objects in this search are the two leading leptons in an event, and the missing tranverse momentum.
Secondary objects used in the selection are jets and additional leptons.
The inner silicon tracker determines trajectories, momenta and vertices of charged particles.
Lead Tungstate electromagnetic calorimeter measures energy and location of electrons and photons.
Brass scintillator hadronic calorimeter measures the energy of jets.
Muon chambers measure the location and momentum of muons.
The information from the subdetectors can be combined to measure the missing transverse momentum in an event.
The scalar magnitude of the missing transverse momentum is referred to as missing transverse energy, or $\MET$.

\section{Invisible Higgs and ZZ Production}

\begin{figure}[h]
\centering
\includegraphics[height=2in]{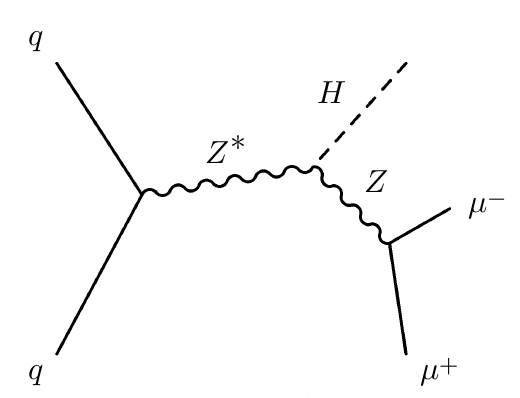}\includegraphics[height=2in]{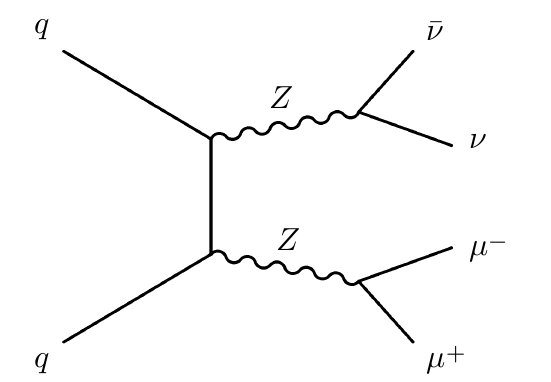}
\caption{(Left) ZH Signal, (Right) ZZ Main background.}
\label{fig:feyn}
\end{figure}

The signal of this search is a Z boson Higgs-strahlung, where the radiated Higgs boson decays invisibly after recoiling from the Z (Fig.~\ref{fig:feyn}, left).
The expected rate of invisible Higgs decay in the standard model is very small~\cite{smdecay}.
Observation of a significant invisible branching fraction would be an indication of new physics.
Since the decay products are not detected, the search is largely independent of the type of invisible decay products from the Higgs boson.
Possible invisible decay modes beyond the standard model include Higgs to dark matter, where the dark matter particles are a pair of lightest supersymmetric particles, like neutralinos~\cite{neutralinos}.
Another theory involves extra dimensions, where the Higgs oscillates into a graviscalar, and then disappears from our brane~\cite{graviscalars}.
The Higgs could also decay into a pair of graviscalars~\cite{graviscalarpairs} or a pair of neutrinos, one light and one heavy~\cite{neutrinos}.

The main background of the search is standard model ZZ production, where one Z boson decays into charged leptons, and the other decays into a pair of neutrinos (Fig.~\ref{fig:feyn}, right). 
It has the same detector signature as the signal: two leptons and missing transverse energy. 
There are some kinematic differences to be expected, since the spin states between ZZ and ZH are different, and there is a mass difference between the Higgs and Z bosons. 
This irreducible ZZ contribution comprises approximately 70\% of the backgrounds remaining at final selection.
\section{Samples and Modeling}


To simulate the signals and backgrounds, a set of monte carlo generators are used.
For ZH (all higgs masses), t$\bar{\text{t}}$, and tW processes, $\POWHEG$ (v2.0)~\cite{powheg} is used.
$\MADGRAPH$ (v5.1.3)~\cite{madgraph} is used to generate diboson and Z$+$jets processes.
For all processes, parton showering is simulated using $\PYTHIA$6 (v4.22 for 7 TeV, v4.26 for 8 TeV)~\cite{pythia}.
The detector response is modeled with $\GEANTfour$~\cite{geant4}.
The parton distribution functions are modeled through CTEQ6L parameterization at leading order, and CT10 parameterization at next to leading order.

Next to leading order cross sections of ZZ and W$^\pm$Z processes are computed using $\MCFM$.
The ZH cross section is computed at NNLO in QCD scale and NLO at electroweak scale, by the LHC Higgs Cross Section Working Group~\cite{LHCWG}.

Data are used to estimate non-resonant backgrounds within the Z boson mass-peak region.
Non-resonant backgrounds include W$^\pm$W$^\mp$, t$\bar{\text{t}}$, tW, single top, and Z$\rightarrow\tau^{+}\tau^{-}$ production.
Scale factors (Eq.~\ref{topeqs}) are calculated from the Z-peak side bands, $40 < m_{\ell\ell} < 70$ GeV and $110 < m_{\ell\ell} < 200$ GeV, of orthogonal ee, $\mu\mu$, and e$\mu$ data control samples.
The scale factors are then applied to the Z-peak region of the e$\mu$ sample to estimate the yields of ee and $\mu\mu$ events:
\begin{eqnarray}
\label{topeqs}
\alpha_{\ell\ell} &=& \frac{N_{\ell\ell}^{\mathrm{SB}}}{N_{\text{e}\mu}^{\mathrm{SB}}} \\
N_{\ell\ell}^{\mathrm{peak}} &=& \alpha_{\ell\ell} \cdot N_{\text{e}\mu}^{\mathrm{peak}} \nonumber \\
\ell &\in& \{\text{e},\mu\} \nonumber
\end{eqnarray}

The resulting scale factors are $\alpha_{\mu\mu} \sim 0.6$ and $\alpha_{\text{e}\text{e}} \sim 0.4$.
This method is checked with a closure test.
A 25\% uncertainty is assigned to this method by comparing the scale factors with ones calculated without a veto on jets tagged as originating from bottom quarks (``b-tagged'').

The Monte Carlo simulation of Z$+$jets may not fully model detector and pile-up effects in the tails of missing energy-type distributions.
To improve modeling of these distributions, the Z$+$jets background is estimated from an orthogonal $\gamma+$jets control sample.
The $\gamma+$jets sample is normalized to match the observed Z$+$jets rate.
Reweighting factors are computed as a function of $p_{\mathrm{T}}^\text{Z}$ and number of reconstructed vertices.
The reweighting by $p_{\mathrm{T}}^\text{Z}$ accounts for dependence of $\MET$ variables on associated hadronic activity.
The reweighting by number of vertices eliminates any discrepencies from effective pile-up, photon trigger prescaling, and event selection.
Electroweak processes with photons and neutrinos are substracted out, using Monte Carlo simulation.
An 100\% uncertainty is assigned to this method, but the absolute contribution after final selection is small.

\section{The Search Strategy}

Events from $\ZHexpl$ and $\ZZexpl$ are characterized by large $\MET$ from neutrinos or non-standard Higgs decay products.
The large potential background, prior to final selection, comes from Z$+$jets events with large mis-measured $\MET$ from hadronic recoil.
This can flood the signal region with Z$+$jets events since the cross section of Z$+$jets is five orders of magnitude greater than that of $\ZZexpl$.

\begin{figure}[h]
\centering
\includegraphics[height=2.3in]{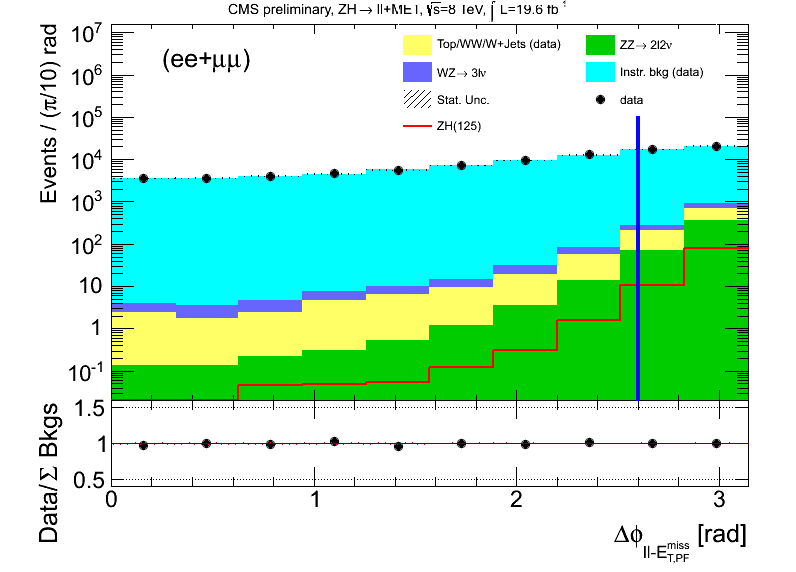}\includegraphics[height=2.3in]{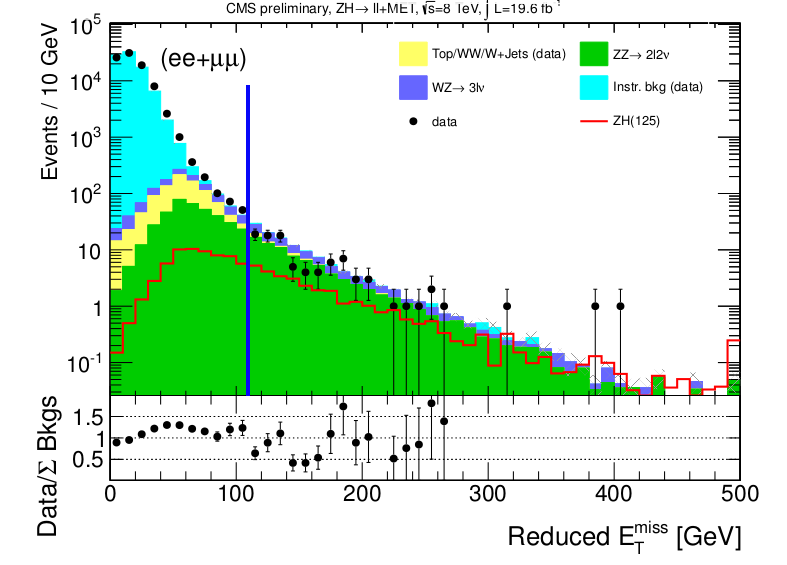}
\caption{(Left) Angle between $\vec{p_{\mathrm{T}}^{\ell\ell}}$ and $\vec{\MET}$, (Right) Reduced Missing Energy; both are shown with optimal selection.}
\label{fig:redmet}
\end{figure}

To reduce the number of mis-measured energy events in the signal, a variable called ``reduced missing transverse energy'' ($\RMET$) is constructed using a similar approach to that used in similar analyses at \DZERO~\cite{DZero}~\cite{DZero2} and OPAL~\cite{OPAL}. 
Two types of energy vectors are designated, ``clustered" energy sums hadronic activity (Eq.~\ref{eq:cl}), and ``unclustered" energy sums non-hadronic activity (Eq.~\ref{eq:uncl}).
The basis of the vectors is defined such that one axis is perpendicular to $\vec{ p_{\mathrm{T}}^{\ell\ell}}$  and the other is parallel.
The energy vectors are projected onto these axes, represented by index $i$ (Eq.~\ref{eq:i}), resulting in a set of two components for each energy vector.
Jet energy and $\MET$ resolution dominate along the perpendicular axis, and real missing energy is significant along the parallel axis.
So the two energy types are minimized by component (Eq.~\ref{eq:rmet}), and added to $\vec{ p_{\mathrm{T}}^{\ell\ell}}$  to make the reduced missing energy, $\RMET$, variable (Eq.~\ref{eq:armet}), (Fig.~\ref{fig:redmet}, right):
\begin{eqnarray}
\label{meteqs}
R_{\mathrm{cl}}^i &=& \sum \limits_{\mathrm{jet}}^{N_{\mathrm{jets}}} p_{\mathrm{T}}^{\mathrm{jet},i} \label{eq:cl}\\
R_{\mathrm{uncl}}^i &=& -\MET^i - p_{\mathrm{T}}^{\ell\ell,i} \label{eq:uncl}\\
\RMET^i &=& p_{\mathrm{T}}^{\ell\ell,i} + \mathrm{min}(R_{\mathrm{cl}}^i,R_{\mathrm{uncl}}^i) \label{eq:rmet}\\
\RMET^2 &=& \sum \limits_{i} (\RMET^i)^2 \label{eq:armet} \\
i &\in& \{\perp (\vec{p_{\mathrm{T}}^{\ell\ell}}), \parallel (\vec{p_{\mathrm{T}}^{\ell\ell}})\} \label{eq:i}
\end{eqnarray}

The reason for preference of $\RMET$ over $\MET$ is that $\RMET$ performs better in signal efficiency and Drell-Yan background suppression. 
Signal efficiency has a $\sim20\%$ relative improvement between $\MET$ and $\RMET$ at the working point.
The $\RMET$ variable is also found to be more stable under pile-up condition and jet energy scale variations by $\sim40\%$.

The selection on events is grouped into two categories, one being the set of main selection and the other being the set of optimized selection.
The main selection focuses on isolating events with a Z boson decaying into leptons.
Events are selected to have two, well-identified, isolated, and same-flavor leptons.
The tranverse momentum of each lepton is required to be above 20 GeV.
The invariant mass of the two leptons is required to be within the Z mass window, $|m_{\ell\ell} - m_\text{Z}|< 15$ GeV.
Events are rejected if there are additional leptons of $p_{\mathrm{T}} > 10$ GeV, to supress W$^\pm$Z events.
Since the signal is not characterized by any hadronic activity, events with large jets, $E_{\mathrm{T}} > 30$ GeV, are rejected.
Also, if a jet is btagged, has $p_{\mathrm{T}} > 20$ GeV and $|\eta| < 2.5$, or if there is a soft muon with $p_{\mathrm{T}} > 3$ GeV, the event is rejected.

The optimized selection is performed on a set of three variables, the angle between $\vec{ p_{\mathrm{T}}^{\ell\ell}}$  and $\vec{\MET}$, the ratio between $\MET$ and $p_{\mathrm{T}}^{\ell\ell}$, and $\RMET$.
All three are optimized together to obtain the best expected exclusion limit at 95\% confidence level, treating ZH with a Higgs mass of 125 GeV as signal. 
The resulting selection is: $\Delta \phi_{\ell\ell \text{-} \METL} > 2.6$ (Fig.~\ref{fig:redmet}, left), $0.8< \frac{\METL}{p_{\mathrm{T}}^{\ell\ell}}<1.2$, and $\RMET > 110$ GeV (Fig.~\ref{fig:redmet}, right). 
This optimized selection is applied to the other ZH samples of different Higgs masses.

\section{Uncertainties}
Uncertainties on the efficiencies of ZH, ZZ, and W$^\pm$Z signals are derived from varying QCD scale, $\alpS$, and PDF type.
Since a shape analysis is performed for computing the limit, errors are propagated to the rate and shape, listed in Table~\ref{tab:unc}.
The combined relative uncertainty on signal efficiency is 12\%, and is dominated by theoretical and PDF uncertainties.
Total relative background estimation uncertainty is 15\%, and is dominated by theoretical uncertainties of W$^\pm$Z and ZZ.
\begin{table}[h]
\begin{small}
\begin{center}
\begin{tabular}{|l|c|c|} \hline
					Type & Source & Uncertainty($\%$)\\ \hline
					 & PDF & 4-5\\
					 & QCD scale variation (ZH) & 7\\
					Rate & QCD scale variation (VV) & 7-10\\
					 & Luminosity & 2.2-4.4\\
					 & Lepton Trigger, Reco., Iso. & 3\\
					 & Z$/\gamma* \rightarrow \ell\ell$ normalization & 100\\
					 & Top, W$^\pm$W$^\mp$, W$^\pm+$jets normalization & 25-100\\ \hline
					 & MC statistics ZH,ZZ,W$^\pm$Z & 1-5\\
					 & Control sample statistics Z$/\gamma* \rightarrow \ell\ell$ & 12-24\\
					Shape & Control sample statistics NRB & 53-100\\
					 and & Pile-up & 0.1-0.3\\
					Rate & b-tagging Efficiency & 0.2\\
					 & Lepton Momentum Scale & 1\\
					 & Jet Energy Scale, Resolution & 1-3\\
					 & Unclustered energy & 1-4\\ \hline
\end{tabular}
\caption{Systematic Uncertainties of ZH Analysis.}
\label{tab:unc}
\end{center}
\end{small}
\end{table}

\section{Results}

\begin{figure}[h]
\centering
\includegraphics[height=2in]{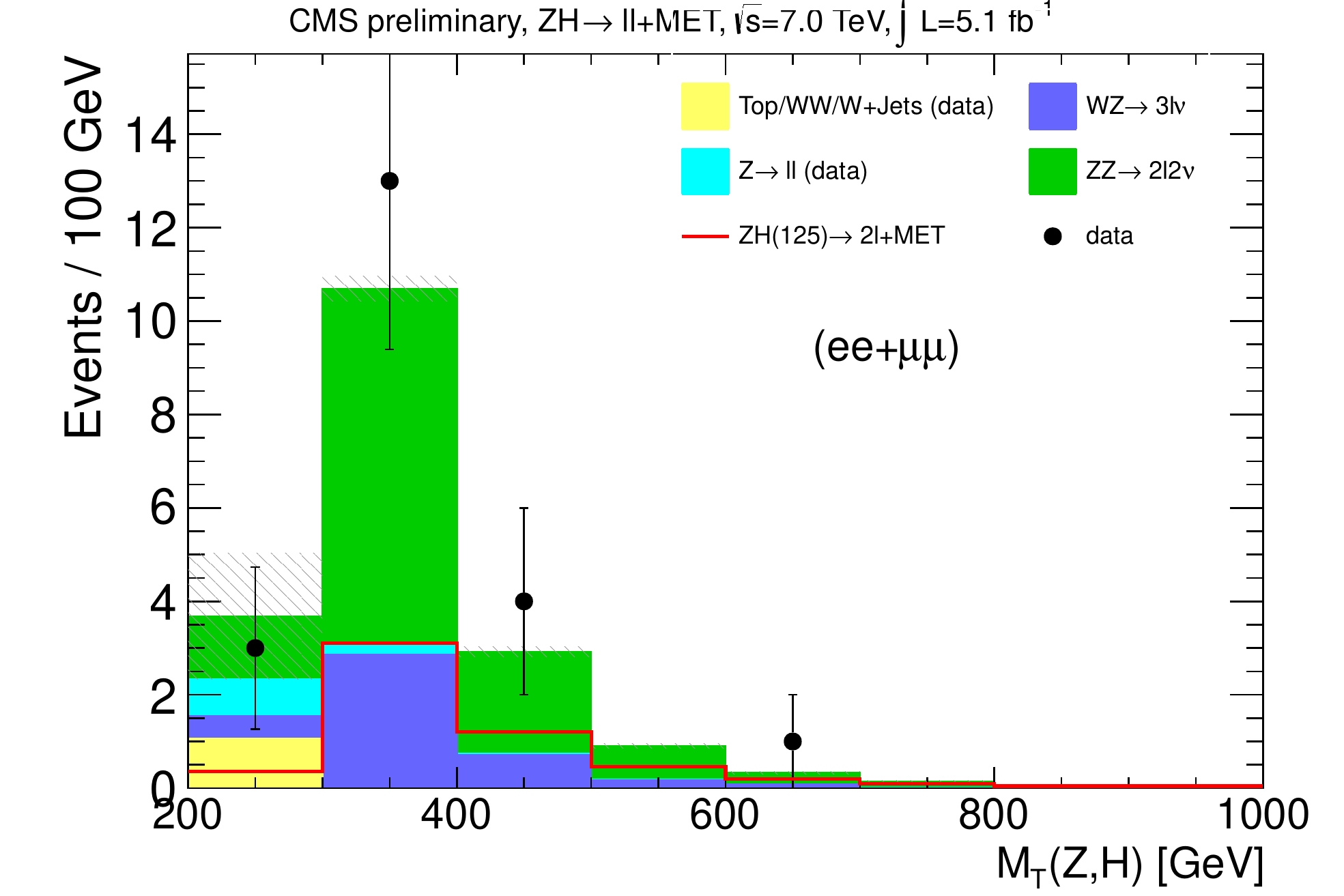}\includegraphics[height=2in]{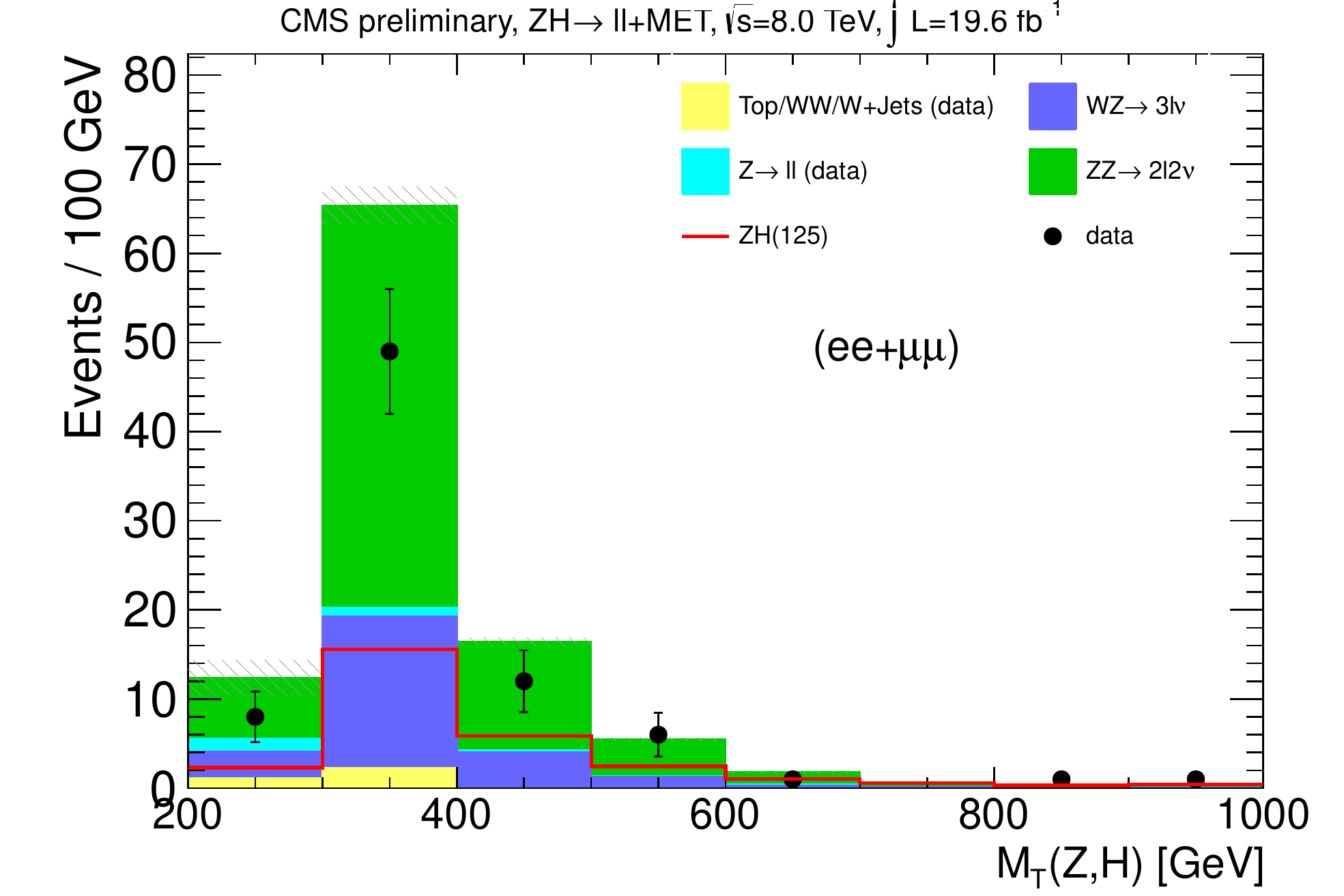}
\caption{(Left) Transverse Mass for 7 TeV, (Right) Transverse Mass for 8 TeV.}
\label{fig:mt}
\end{figure}

At final selection, no significant excess is observed (Table~\ref{tab:yields}). 
A shape analysis is performed on the $m_{\mathrm{T}}$ distribution, which is a ``pseudo tranverse mass" (Fig.~\ref{fig:mt}) between the Z boson and the missing energy.
It is defined as:
\begin{equation}
\label{mteq} 
m_{\mathrm{T}}^2 = \left(\sqrt{ {p_{\mathrm{T}}^{\ell\ell}}^2 + m_{\ell\ell}^2 } + \sqrt{ {\MET}^2 + m_{\ell\ell}^2 }\right)^2 - \left({\vec{p_{\mathrm{T}}^{\ell\ell}}} + \vec{\MET}\right)^2.
\end{equation}
The Z mass is used as a placeholder for the mass of the invisibly decaying particle.
This variable exploits the kinematic differences between $\ZHexpl$ and $\ZZexpl$ to improve the limit.
Both signals have missing energy, but the mass and spin of the missing particle is different.
The shape analysis is done for all Higgs masses.
\begin{table}[h]
\begin{small}
\begin{center}
\begin{tabular}{|l|c|c|c|c|} \hline
	& \multicolumn{2}{c|}{$\sqrt{s} = 7$ TeV} & \multicolumn{2}{c|}{$\sqrt{s} = 8$ TeV} \\
	Process & ee & $\mu\mu$ & ee & $\mu\mu$ \\ \hline
	\textcolor{darkgreen}{ZH$(m_\text{H}=125\text{ GeV})$} & \textcolor{darkgreen}{$2.2\pm0.3$} & \textcolor{darkgreen}{$3.3\pm0.5$} & \textcolor{darkgreen}{$11.8\pm1.9$} & \textcolor{darkgreen}{$16.7\pm2.5$} \\ \hline
	Z$/\gamma* \rightarrow \ell\ell$ & $0.3\pm0.3$ & $0.7\pm0.7$ & $1.0\pm1.0$ & $1.9\pm1.9$ \\
	Top/W$^\pm$W$^\mp$/W$^\pm+$jets & $0.4\pm0.4$ & $0.6\pm0.6$ & $1.3\pm0.8$ & $2.1\pm1.3$ \\
	W$^\pm$Z$\rightarrow 3\ell \cPgn$ & $2.0\pm0.3$ & $2.3\pm0.3$ & $11.0\pm1.6$ & $14.8\pm2.1$ \\
	$\ZZexpl$ & $5.1\pm0.6$ & $7.3\pm0.8$ & $29.8\pm3.6$ & $40.8\pm4.5$ \\ \hline
	\textcolor{blue}{total bkgd} & \textcolor{blue}{$7.8\pm0.8$} & \textcolor{blue}{$11.0\pm1.3$} & \textcolor{blue}{$43.1\pm4.1 $} & \textcolor{blue}{$59.6\pm5.5$} \\ \hline
	 Data & 10 & 11 & 33 & 45 \\ \hline
\end{tabular}
\caption{Final Yields of ZH Analysis, 100\% $BR($H$\rightarrow$invisible$)$ is assumed in computing signal yields.}
\label{tab:yields}
\end{center}
\end{small}
\end{table}

\begin{figure}[h]
\centering
\includegraphics[height=2in]{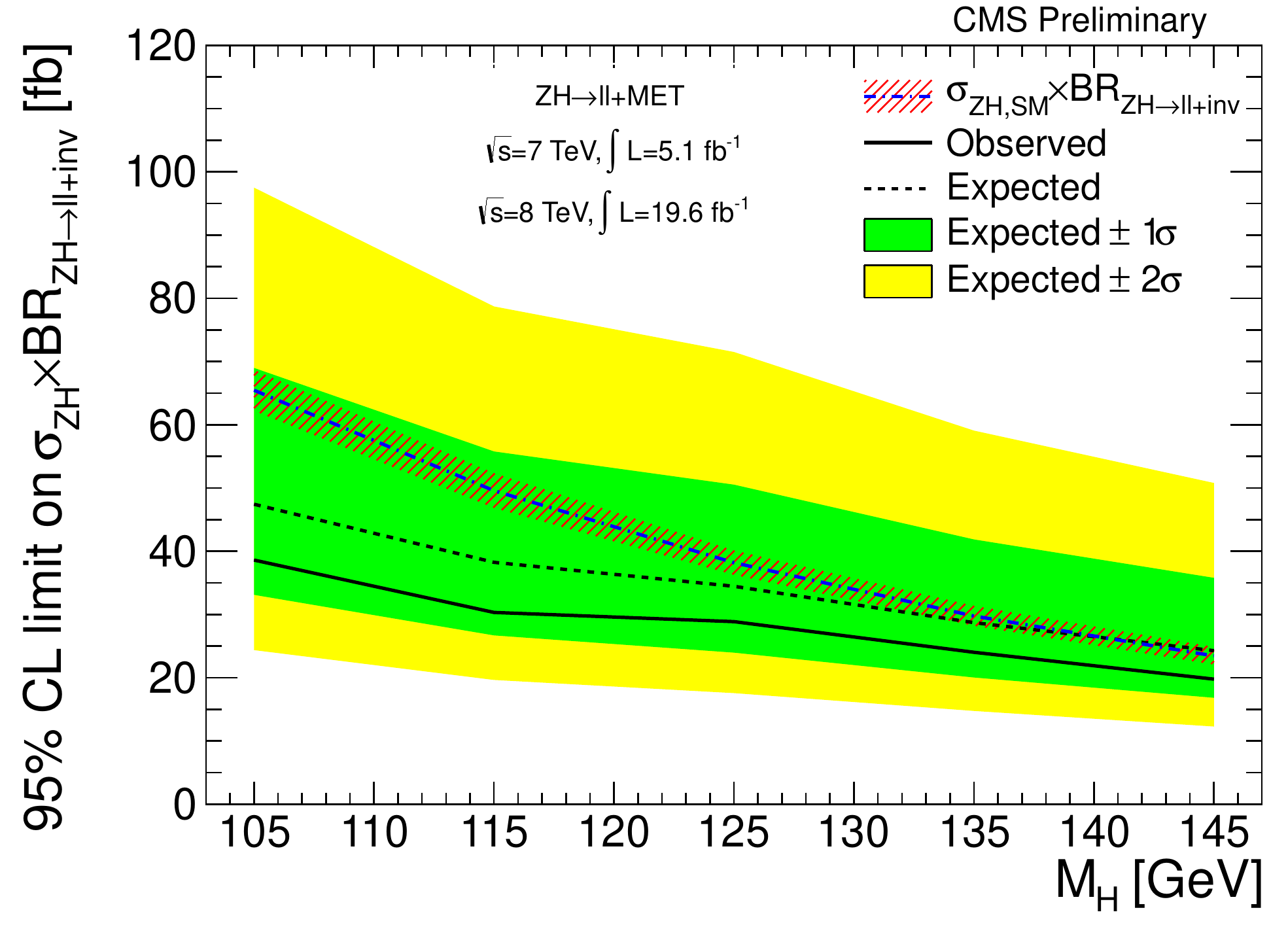}\includegraphics[height=2in]{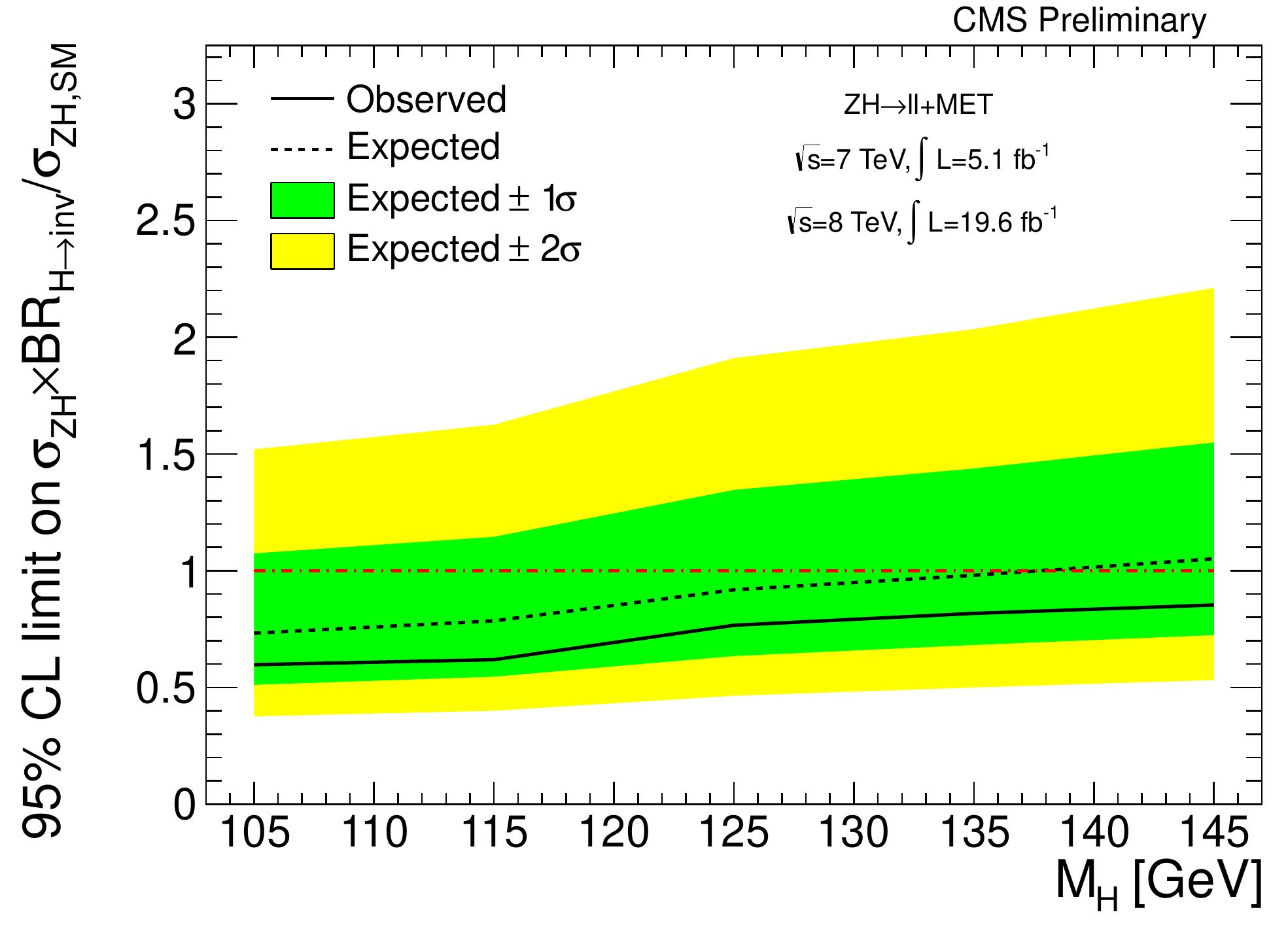}
\caption{(Left) Limit on $\sigma_{\text{ZH}} \cdot BR_{\text{Z}\rightarrow\ell\ell} \cdot BR_{\text{H}\rightarrow \text{inv}}$, (Right) Limit on $BR_{\text{H}\rightarrow \text{inv}}$; both are with 7 and 8 TeV samples combined.}
\label{fig:limits}
\end{figure}

Using a modified frequentist construction $CL_S$, with profile-likelihood test statistics, an upper limit is set on the invisible Higgs production cross section (Fig.~\ref{fig:limits}, left).
Log-normal prior probability are used to describe systematic uncertainties.
A binned shape analysis is performed using the $m_{\mathrm{T}}$ distribution after final selection for each mass point.
The variables $p_{\mathrm{T}}^\text{Z}$ and $\MET$ were considered as well, but $m_{\mathrm{T}}$ has lowest expected limit for each mass point.
If a standard model production rate is assumed, the limit can be expressed as a limit on the branching fraction (Fig.~\ref{fig:limits}, right).
For a Higgs of mass 125 GeV, the expected 95\% C.L. upper limit on $BR($H$\rightarrow$invisible$)$ is 91\%, the observed is 75\%. The limits corresponding to all Higgs masses are listed in Table~\ref{tab:limits}.
\begin{table}[h]
\begin{small}
\begin{center}
\begin{tabular}{|l|c|c|c|c|c|} \hline
					$m_\text{H}$ (GeV) & 105 & 115 & 125 & 135 & 145 \\ \hline
					Obs Lim($\%$) & 60 & 63 & 75 & 82 & 85 \\
					Exp Lim($\%$) & 73 & 79 & 91 & 97 & 105 \\ \hline
\end{tabular}
\caption{Limits on $BR($H$\rightarrow$invisible$)$ for Each Higgs Mass.}
\label{tab:limits}
\end{center}
\end{small}
\end{table}

\section{Conclusion}
The direct search for invisible decay products of a Higgs boson at the LHC, using the CMS experiment, is performed for a standard model-like Higgs radiated through a Z Higgs-strahlung. 
The study uses the full 7 and 8 TeV data samples. 
The mass range explored is 105-145 GeV, and no excess is observed in data.
The results can be interpreted as a limit on the invisible branching fraction, assuming a standard model production rate.
For a Higgs boson with mass of 125 GeV, the observed (expected) 95\% C.L. upper limit on $BR($H$\rightarrow$invisible$)$ is 75\% (91\%).


\end{document}

%% file: econfmacros.tex
\def\beq{\begin{equation}}
\def\eeq#1{\label{#1}\end{equation}}
\def\eeqn{\end{equation}}

\def\beqa{\begin{eqnarray}}
\def\eeqa#1{\label{#1}\end{eqnarray}}
\def\eeqan{\end{eqnarray}}

\let\bar=\overbar

\def\Dslash{\not{\hbox{\kern-4pt $D$}}}
\def\dslash{\not{\hbox{\kern-2pt $\del$}}}

\def\msb{{\bar{\ssstyle M \kern -1pt S}}}




%% file: CMSinvH.bbl
\begin{thebibliography}{99}


\bibitem{CMSind} 
  [CMS Collaboration],
  CMS-PAS-HIG-13-005.

\bibitem{ATLASind} 
  [ATLAS Collaboration],
  ATLAS-CONF-2013-034.

\bibitem{ATLASdir} 
  [ATLAS Collaboration],
  ATLAS-CONF-2013-011.

\bibitem{CMS} 
  S.~Chatrchyan {\it et al.}  [CMS Collaboration],
  JINST {\bf 3}, S08004 (2008).

\bibitem{smdecay}
  A.~Denner, S.~Heinemeyer, I.~Puljak, D.~Rebuzzi and M.~Spira,
  Eur.\ Phys.\ J.\ C {\bf 71}, 1753 (2011)
  [arXiv:1107.5909 [hep-ph]].

\bibitem{neutralinos}
  J.~Beringer {\it et al.}  [Particle Data Group Collaboration],
  Phys.\ Rev.\ D {\bf 86}, 010001 (2012).

\bibitem{graviscalars} 
  G.~F.~Giudice, R.~Rattazzi and J.~D.~Wells,
  Nucl.\ Phys.\ B {\bf 595}, 250 (2001)
  [hep-ph/0002178].

\bibitem{graviscalarpairs} 
  M.~Battaglia, D.~Dominici, J.~F.~Gunion and J.~D.~Wells,
  hep-ph/0402062.

\bibitem{neutrinos} 
  J.~-H.~Chen, X.~-G.~He, J.~Tandean and L.~-H.~Tsai,
  Phys.\ Rev.\ D {\bf 81}, 113004 (2010)
  [arXiv:1001.5215 [hep-ph]].

\bibitem{powheg} 
  S.~Alioli, P.~Nason, C.~Oleari and E.~Re,
  JHEP {\bf 1006}, 043 (2010)
  [arXiv:1002.2581 [hep-ph]].

\bibitem{madgraph} 
  J.~Alwall, M.~Herquet, F.~Maltoni, O.~Mattelaer and T.~Stelzer,
  JHEP {\bf 1106}, 128 (2011)
  [arXiv:1106.0522 [hep-ph]].

\bibitem{pythia}
  T.~Sjostrand, S.~Mrenna and P.~Z.~Skands,
  JHEP {\bf 0605}, 026 (2006)
  [hep-ph/0603175].

\bibitem{geant4} 
  S.~Agostinelli {\it et al.}  [GEANT4 Collaboration],
  Nucl.\ Instrum.\ Meth.\ A {\bf 506}, 250 (2003).

\bibitem{LHCWG}
  S.~Dittmaier {\it et al.}  [LHC Higgs Cross Section Working Group Collaboration],
  arXiv:1101.0593 [hep-ph].

\bibitem{DZero} 
  V.~M.~Abazov {\it et al.}  [D0 Collaboration],
  Phys.\ Rev.\ D {\bf 78}, 072002 (2008)
  [arXiv:0808.0269 [hep-ex]].

\bibitem{DZero2} 
  V.~M.~Abazov {\it et al.}  [D0 Collaboration],
  Phys.\ Rev.\ D {\bf 85}, 112005 (2012)
  [arXiv:1201.5652 [hep-ex]].

\bibitem{OPAL} 
  K.~Ackerstaff {\it et al.}  [OPAL Collaboration],
  Eur.\ Phys.\ J.\ C {\bf 4}, 47 (1998)
  [hep-ex/9710010].

\bibitem{CMSdir}
  [CMS Collaboration],
  CMS-PAS-HIG-13-018.

\end{thebibliography}
